\documentclass[12pt,preprint]{emulateapj}

\def\lxi{\log_{10} \xi_{ion}/[\textrm{Hz ergs}^{-1}]}
\def\lxiz{\log_{10} \xi_{ion,0}/[\textrm{Hz ergs}^{-1}]}
\def\lxif{\log_{10} f_{esc}\xi_{ion}/[\textrm{Hz ergs}^{-1}]}

\begin{document}
\title{The Lyman-Continuum Photon Production Efficiency $\xi_{ion}$ of
  $z\sim4$-5 Galaxies from IRAC-based H$\alpha$ Measurements:
  Implications for the Escape Fraction and Cosmic Reionization}
\author{R.J. Bouwens\altaffilmark{1}, R. Smit\altaffilmark{1,2},
  I. Labb{\'e}\altaffilmark{1}, M. Franx\altaffilmark{1},
  J. Caruana\altaffilmark{3}, P. Oesch\altaffilmark{3},
  M. Stefanon\altaffilmark{1}, N. Rasappu\altaffilmark{1}}
\altaffiltext{1}{Leiden Observatory, Leiden University, NL-2300 RA
  Leiden, Netherlands} \altaffiltext{2}{Centre for Extragalactic
  Astronomy, Durham University, South Road, Durham, DH1 3LE, UK}
\altaffiltext{3}{Institute of Space Sciences \& Astronomy, University
  of Malta, Msida MSD 2080, Malta}
\altaffiltext{4}{Yale Center for
  Astronomy and Astrophysics, Yale University, New Haven, CT 06520,
  USA}

\begin{abstract}
Galaxies represent one of the preferred candidate sources to drive the
reionization of the universe.  Even as gains are made in mapping the
galaxy $UV$ luminosity density to $z>6$, significant uncertainties
remain regarding the conversion to the implied ionizing emissivity.
The relevant unknowns are the Lyman-continuum (LyC) photon production
efficiency $\xi_{ion}$ and the escape fraction $f_{esc}$.  As we show
here, the first of these unknowns is directly measureable in $z=4$-5
galaxies based on the impact the $H\alpha$ line has on the observed
IRAC fluxes.  By computing a LyC photon production rate from the
implied H$\alpha$ luminosities for a broad selection of $z=4$-5
galaxies and comparing this against the dust-corrected $UV$-continuum
luminosities, we provide the first-ever direct estimates of the LyC
photon production efficiency $\xi_{ion}$ for the $z\geq4$ galaxy
population.  We find $\lxi$ to have a mean value of
$25.27_{-0.03}^{+0.03}$ and $25.34_{-0.02}^{+0.02}$ for sub-$L^*$
$z=4$-5 galaxies adopting Calzetti and SMC dust laws, respectively.
Reassuringly, both derived values are consistent with standardly
assumed $\xi_{ion}$'s in reionization models, with a slight preference
for higher $\xi_{ion}$'s (by $\sim$0.1 dex) adopting the SMC dust law.
High values of $\xi_{ion}$ ($\sim$25.5-25.8 dex) are derived for the
bluest galaxies ($\beta<-2.3$) in our samples, independent of dust law
and consistent with results for a $z=7.045$ galaxy.  Such elevated
values of $\xi_{ion}$ would have important consequences, indicating
that $f_{esc}$ cannot be in excess of 13\% for standard assumptions
about the faint-end cut-off to the LF and the clumping factor.
\end{abstract}

\section{Introduction}

One of the biggest longstanding puzzles regards the reionization of
the universe.  While we have general knowledge of the broad time scale
over which reionization has occurred, many important issues remain
unclear.  For example, there continues to be a debate about which
sources drive reionization (e.g., Robertson et al.\ 2015; Madau \&
Haardt 2015).  Similarly, we have limited information about the
precise epoch when reionization is completed and also how rapidly the
universe transitions from a largely neutral state to the $\sim$30\%
ionized filling factors being inferred at $z\sim8$ (Schenker et
al.\ 2014; Robertson et al.\ 2015; Bouwens et al.\ 2015b; Mitra et
al.\ 2015; Ishigaki et al.\ 2015; Finkelstein 2015).

In the last year, new estimates of the Thomson optical depths
($\tau=0.066\pm0.016$) have become available thanks to an analysis of
the results from the Planck mission (Planck Collaboration 2015) and
are consistent with the cosmic ionizing emissivity being somewhat
lower than what had previously been inferred from analyses of the WMAP
$\tau$ measurements (e.g., Kuhlen \& Faucher-Gigu{\`e}re 2012; Haardt
\& Madau 2012; Bouwens et al.\ 2012a; Alvarez et al.\ 2012; Robertson
et al.\ 2013).  These new results point towards the cosmic ionizing
emissivity evolving very similarly to the $UV$-continuum luminosity
density (Robertson et al.\ 2015; Bouwens et al.\ 2015b; Mitra et
al.\ 2015; Choudhury et al.\ 2015).

In calculating the ionizing emissivity derived from galaxies, three
factors are standardly included in the calculation (e.g., Kuhlen \&
Faucher-Gigu{\`e}re 2012; Robertson et al.\ 2013): the galaxy $UV$
luminosity density $\rho_{UV}$, the escape fraction $f_{esc}$, and the
Lyman-continuum photon production efficiency $\xi_{ion}$ (describing
the production rate of Lyman-continuum ionizing photons per unit
luminosity in the $UV$-continuum).  While most of the effort has been
devoted to improving current constraints on the $UV$ luminosity
density $\rho_{UV}$ and the escape fraction $f_{esc}$, the
Lyman-continuum photon production efficiency $\xi_{ion}$ is also
fairly uncertain.  In general, estimates of this efficiency
$\xi_{ion}$ appear to be exclusively indirect, based on the
$UV$-continuum slope $\beta$ of galaxies using standard stellar
population models (Robertson et al.\ 2013; Duncan \& Conselice 2015;
Bouwens et al.\ 2015b, 2015c) or using predictions for young stellar
populations that are possibly subsolar (e.g., Madau et al.\ 1999;
Schaerer 2003).

Despite these indirect attempts to constrain $\xi_{ion}$, many recent
observations are now providing constraints on the H$\alpha$ fluxes of
$z\sim4$ and $z\sim5$ galaxies based on the impact of H$\alpha$ and
other nebular lines to the IRAC fluxes (Schaerer \& de Barros et
al.\ 2009; Shim et al.\ 2011; Stark et al.\ 2013; de Barros et
al.\ 2014; Laporte et al.\ 2014; Rasappu et al.\ 2015; Smit et
al. 2015a,b; Marmol-Queralto et al.\ 2015).  As the observed H$\alpha$
fluxes can be directly related to the total number of Lyman-continuum
photons produced by stars in a galaxy (assuming an escape fraction of
zero: e.g., Leitherer \& Heckman 1995), we can use the observed
H$\alpha$ and $UV$-continuum fluxes of distant galaxies to set
constraints on $\xi_{ion}$.  In a related investigation, Stark et
al.\ (2015) recently showed how one could use measurements of the flux
in the CIV$\lambda$1548 line for a lensed Lyman-break galaxy at
$z=7.045$ to constrain $\lxi$, estimating it to be
$25.68_{-0.19}^{+0.27}$.

Here we derive constraints on the Lyman-continuum photon production
efficiency $\xi_{ion}$ by making use of a large sample of star-forming
galaxies distributed over the redshift range $z=3.8$-5.4, where we
know the passband in which the H$\alpha$ emission likely falls.  We
obtain these samples thanks to the recent work of Smit et al.\ (2015b)
and Rasappu et al.\ (2015), where spectroscopic redshift $z=3.8$-5.4
samples are supplemented with photometric redshift samples.  In each
case, the $H\alpha$ emission line lies in one of the two IRAC filters,
with no prominent contribution from other nebular lines.  As we will
see (\S3) and as demonstrated by the results presented in Smit et
al. (2015b), these spectroscopic and photometric samples exhibit
similar H$\alpha$ EWs but have complementary strengths (i.e.,
spectroscopic-redshift samples provide a sampling of galaxies with
more secure redshifts while photometric redshift samples likely
provide a more representative sampling of UV-bright galaxies, with
less bias towards line emitters).

The plan for this paper is as follows.  We begin (\S2) by briefly
summarizing the observational data sets and selection criteria.  In
\S3, we describe the methodology we utilize in Smit et al.\ (2015b)
for deriving the H$\alpha$ fluxes for individual sources in our
different samples.  We then use these H$\alpha$ fluxes to estimate the
Lyman-continuum photon production efficiency $\xi_{ion}$ for
individual sources and then look at how $\xi_{ion}$ depends on the
$UV$ luminosity, the $UV$-continuum slope, and redshift.  We then
combine these measurements with results available on the total
ionizing emissivity at $z\sim4$-5 to set an upper limit on the escape
fraction of galaxies to 13\%.  In \S4, we discuss the implications of
the present results and then conclude (\S5).  Where necessary, we
assume $\Omega_0 = 0.3$, $\Omega_{\Lambda} = 0.7$, and $H_0 =
70\,\textrm{km/s/Mpc}$.  All magnitudes are in the AB system (Oke \&
Gunn 1983).

\section{Observational Data and Sample Selection}

In the present section, we provide a brief summary of both the
observational data sets and selection criteria we utilize for deriving
our results.  As we use the $z=3.8$-5.4 samples and IRAC photometry
from Smit et al.\ (2015b) and Rasappu et al.\ (2015), we refer the
interested reader to those papers for more details.

\subsection{Observation Data}

For our source selection and photometry, we utilize the deep HST
optical and near-infrared observations over the two GOODS fields.
Over those fields, we make use of almost all optical/ACS and
near-infrared/WFC3/IR observations, including observations from the
original ACS GOODS and follow-up program (Giavalisco et al.\ 2004),
the ERS program (Windhorst et al.\ 2011), and the CANDELS program
(Grogin et al.\ 2011; Koekemoer et al.\ 2011).  Collectively, the data
from these programs generally reach to $>$$\sim$27 mag at $5\sigma$
all the way from optical wavelengths at 0.4$\mu$m to the near-infrared
1.6$\mu$m.  Moderately deep observations ($\sim$25.0-25.5 mag:
$5\sigma$) in the $K$-band are available over $>$90\% of the two
CANDELS fields.

For the Spitzer/IRAC observations needed for our H$\alpha$ flux
estimates, we utilize the new reductions from Labb{\'e} et
al.\ (2015), who have incorporated the full set of observations from
the original GOODS, SEDS (Ashby et al.\ 2013), S-CANDELS (Ashby et
al.\ 2015), and IUDF programs (Labb{e et al.\ 2015).  These reductions
  feature a PSF with a 1.8$''$ FWHM, $\sim$10\% sharper than achieved
  in most analyses, due to the use of a drizzle methodology for
  coadding the Spitzer/IRAC observations.

\subsection{Selection of Spectroscopic Sample}

A large number of spectroscopic redshifts have been derived over the
GOODS-North and South over the last ten years and made public in many
independent efforts (Vanzella et al.\ 2005, 2006, 2008, 2009; Balestra
et al.\ 2010; Shim et al.\ 2011; Stark et al.\ 2010, 2011, 2013;
Rasappu et al.\ 2015).

In Smit et al.\ (2015b) and Rasappu et al.\ (2015), we took advantage
of several public spectroscopic redshift compilations (Vanzella et
al.\ 2009; Shim et al.\ 2011; Stark et al.\ 2013) to construct a
sample of $z=3.8$-5.0 galaxies and $z=5.1$-5.4 galaxies, while also
benefitting from some $z=5.1$-5.4 sources from Stark et al.\ (2015, in
prep).  For the first redshift subsample, the H$\alpha$ line falls
squarely in the Spitzer/IRAC $3.6\mu$m filter, and in the second
subsample, the H$\alpha$ line falls in the Spitzer/IRAC $4.5\mu$m
filter.

\subsection{Selection of Photometric Sample}

Following the treatment in Smit et al.\ (2015b) and Rasappu et
al.\ (2015), we also consider a selection of sources which very likely
lie in the redshift intervals $z=3.8$-5.0 and $z=5.1$-5.4,
respectively (99\% and 85\%), according to their photometric
constraints.  Rasappu et al.\ (2015) only required sources to show an
85\% likelihood of lying in the target redshift interval ($z=5.1$-5.4)
to compensate for the greater difficulty of isolating a source
photometrically to such a narrow interval in redshift.

Selecting sources according to their redshift likelihood distribution
is useful, since it allows us to be more inclusive in our selection of
$z\sim4$-5 galaxies and not to base the results on sources which only
show Ly$\alpha$ in emission.  This is to address the concern that such
samples may be biased towards sources with younger ages and not be
totally representative.

\section{Empirical Estimate of $\xi_{ion}$} 

\begin{figure*}
\epsscale{1.1}
\plotone{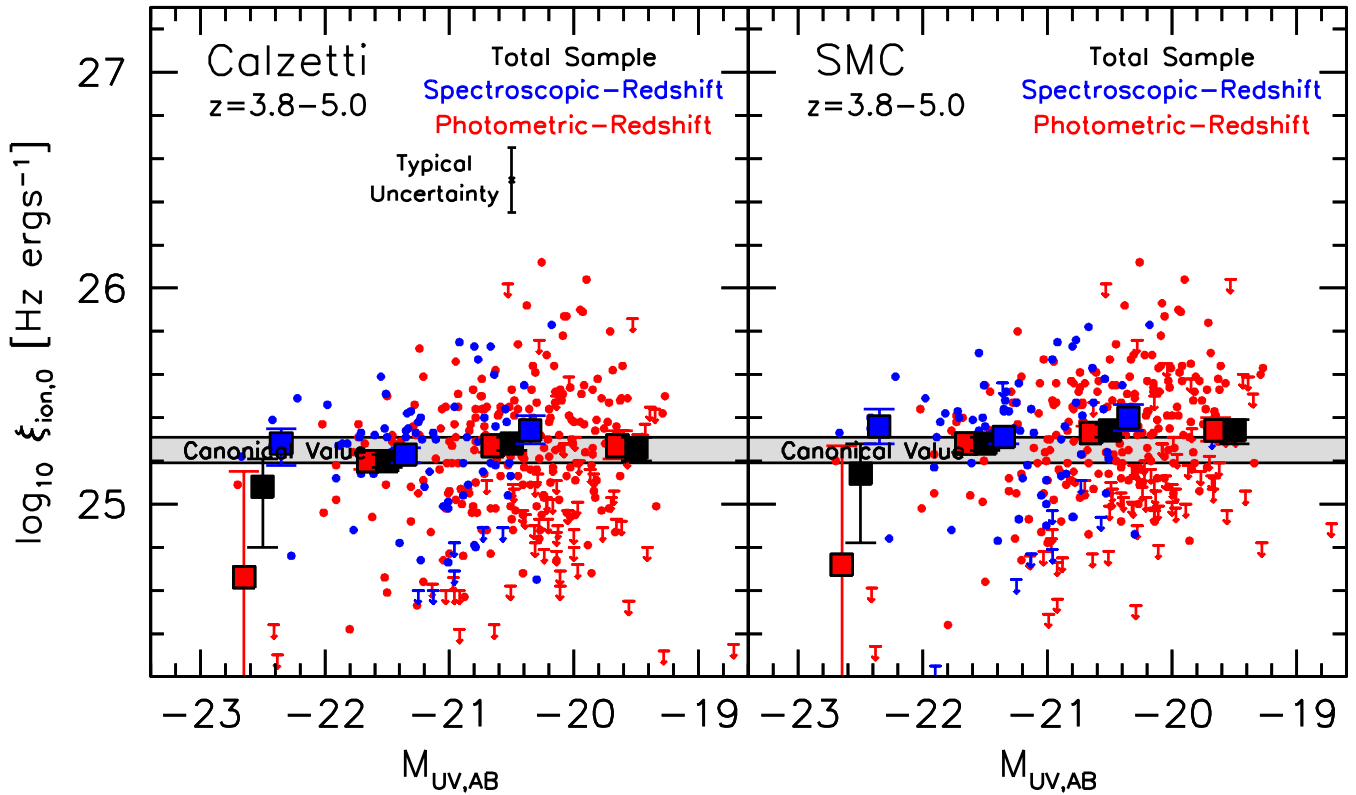}
\caption{Derived Lyman-continuum photon production efficiencies
  $\xi_{ion}$ based on the H$\alpha$ luminosities derived from a fit
  to the IRAC fluxes in $z\sim4$-5 galaxies and assuming a Calzetti et
  al.\ (2000: \textit{left panel}) or SMC-like dust law (\textit{right
    panel}: \S3.2).  A Lyman-continuum escape fraction of zero has
  been assumed in deriving these $\xi_{ion,0}$'s (see \S3.5 for the
  values with non-zero escape fractions).  Sources where spectroscopic
  redshifts or well-determined photometric redshifts place the
  H$\alpha$ line in a specific IRAC band are indicated by the blue and
  red points, respectively.  $1\sigma$ upper limits are included on
  this diagram with downward arrows in cases where the H$\alpha$
  emission line is not detected at $1\sigma$ in the photometry.  The
  solid red and blue squares indicate the mean value of $\xi_{ion}$
  for red and blue colored points, while the solid black square
  indicate the mean values combining the spectroscopic and
  photometric-redshift selected samples (shown for all bins with $>$1
  source and offset from the center of the bin for clarity).  The grey
  band indicates the Lyman-continuum photon production efficiencies
  $\xi_{ion}$ assumed in typical models (Table~\ref{tab:xion_lit}).
  The black error bar near the top of the left panel indicate the
  typical uncertainties in the derived $\xi_{ion}$'s.  The $\xi_{ion}$
  values we observe for both dust laws are consistent with the values
  assumed in canonical reionization models; however, we note a slight
  preference for higher $\xi_{ion}$'s adopting the SMC dust
  law.\label{fig:xi}}
\end{figure*}

\subsection{Measurement of H$\alpha$ Fluxes}

The H$\alpha$ flux measurements we utilize in this study are directly
taken from Smit et al.\ (2015b) and from Rasappu et al.\ (2015), so we
refer our audience to their studies for a detailed description.
Nevertheless, our basic methodology is as follows.  To begin, we
derive a detailed SED fit to the full photometry we have available
(HST + ground-based $K_s$ + Spitzer/IRAC) for all sources in our
samples to obtain good constraints on the overall shape of the
spectral energy distribution, excluding the Spitzer/IRAC passband we
expect with high confidence to contain the H$\alpha$ emission line.
We then compare the measured flux of sources in the Spitzer/IRAC
3.6$\mu$m or 4.5$\mu$m band with the model flux expected in that band
based on our best-fit SED (and not including line flux in the SED
model).

This procedure leads to an estimate of the flux in the H$\alpha$ line
and other lines at approximately the same wavelength.  One such line
is [NII], but other lines (e.g., [SII]) also contribute.  As in Smit
et al.\ (2015b) and Rasappu et al.\ (2015), we estimate the impact of
the [NII] emission line on the measured H$\alpha$ flux based on the
model results of Anders \& Fritze-v.~Alvensleben (2003), where
[NII]/H$\alpha$ is 6.8\% and [SII]/H$\alpha$ is 9.5\%.  These line
ratios are very consistent with that found for normal to lower-mass
galaxies at $z=2.9$-3.8 galaxies (e.g., Sanders et al.\ 2014).  We
verified that the model SED fits for all sources used in our study
were sufficiently good as to produce credible measurement of
$L_{H\alpha}$, and no sources from Smit et al.\ (2016) were excluded.

The present methodology is almost identical to the methodology
employed in Shim et al.\ (2011), Stark et al.\ (2013), and most
recently Marmol-Queralto et al.\ (2015).  While one might be concerned
that this approach may lead to small systematic errors in the fluxes
in various emission lines, one can test the accuracy of the flux
measurements by comparing the [3.6]$-$[4.5] colors of $3.1<z<3.6$,
$z=3.8$-5.0, and $z=5.1$-5.4 galaxy samples.  Encouragingly enough,
the estimated EWs one derives from differential comparisons of the
[3.6]$-$[4.5] colors agree very well with the fit results performed on
the individual SEDs.  For example, in Rasappu et al.\ (2015: comparing
$z=4.4$-5.0 and $z=5.1$-5.4 samples), the mean H$\alpha$ EW we derive
from the SED fits for the photometric samples is 638$\pm$118 \AA$\,$
(vs. 665$\pm$53 \AA$\,$from the differential comparison) and
855$\pm$179 \AA$\,$for the spectroscopic samples (vs. 707$\pm$74
\AA$\,$ from the differential comparison).  Marmol-Queralto et
al.\ (2015) also demonstrate that they achieve equivalent constraints
on the H$\alpha$+[NII] EWs for $z\sim1.3$ galaxies using the HST
WFC3/IR grism observations from the 3D-HST program (Brammer et
al.\ 2012) as they find using the present SED-fitting procedure.

\begin{figure*}
\epsscale{1.1}
\plotone{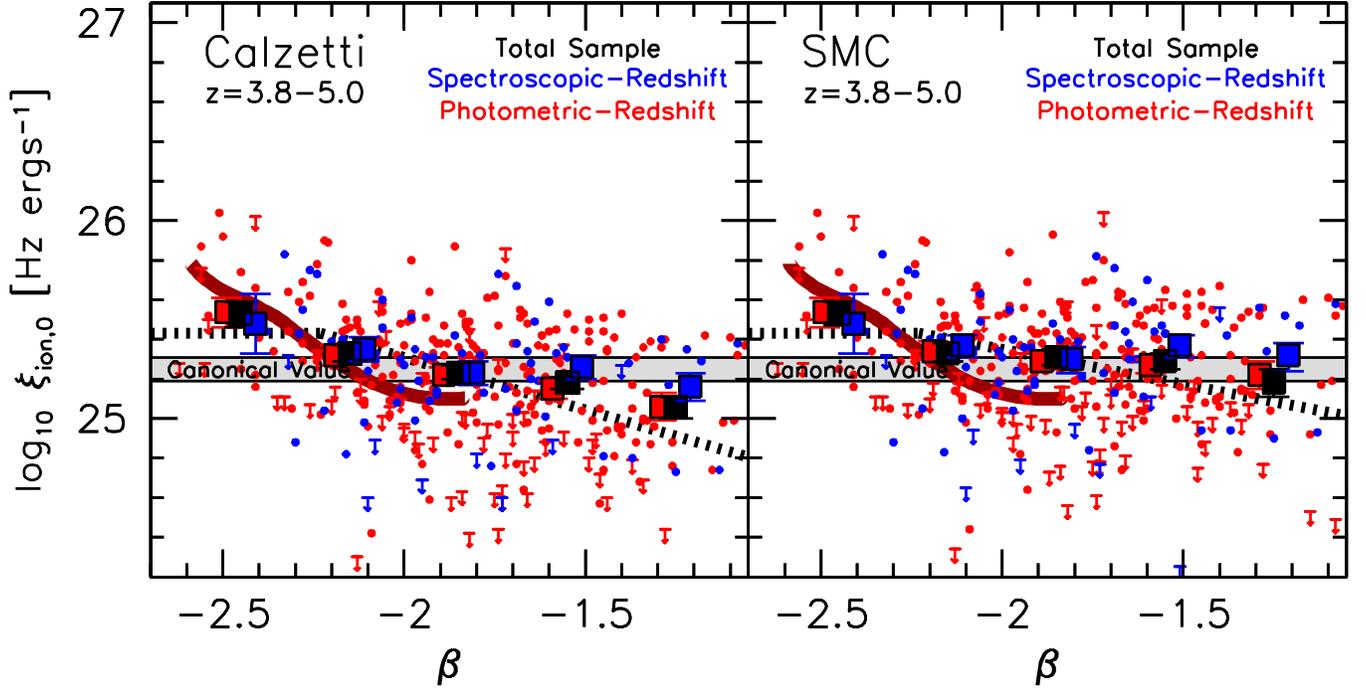}
\caption{Dependence on $\xi_{ion}$'s we have derived on the
  $UV$-continuum slope $\beta$ assuming either a Calzetti et
  al.\ (2000) extinction law (\textit{left panel}) or a SMC-like
  extinction law (\textit{right panel}: \S3.2).  A Lyman-continuum
  escape fraction of zero has been assumed in deriving these
  $\xi_{ion,0}$'s (see \S3.5 for the values with non-zero escape
  fractions).  The red, blue, and black symbols are the same as on
  Figure~\ref{fig:xi}.  The thick dotted lines show the trend in
  $\xi_{ion,0}$ vs. $\beta$ that would result from the impact of dust
  corrections on the observed IRAC excesses and $UV$ magnitudes.  The
  thick red line indicates the predicted $\xi_{ion}$ vs. $\beta$ trend
  for a stellar population model with zero dust extinction, a
  metallicity of 0.4$Z_{\odot}$, and a range in ages using the Bruzual
  \& Charlot (2003) models (see Robertson et al.\ 2013; Duncan \&
  Conselice 2015; Bouwens et al.\ 2015b).  Independent of our
  assumptions about the dust law, we consistently derive higher values
  for $\xi_{ion}$ (by $\sim$0.2 dex) for the bluest galaxies than have
  been canonically assumed for the star-forming population as a whole
  (but consistent with the higher values suggested by Duncan \&
  Conselice 2015 and Bouwens et al.\ 2015b for the bluest galaxies).
  Our $\xi_{ion}$ results for both dust laws are consistent with
  canonically assumed values.  We note a slight preference for higher
  values (by $\sim$0.1 dex) of $\xi_{ion}$ adopting the SMC dust
  law.\label{fig:xibeta}}
\end{figure*}

\begin{figure*}
\epsscale{1.1}
\plotone{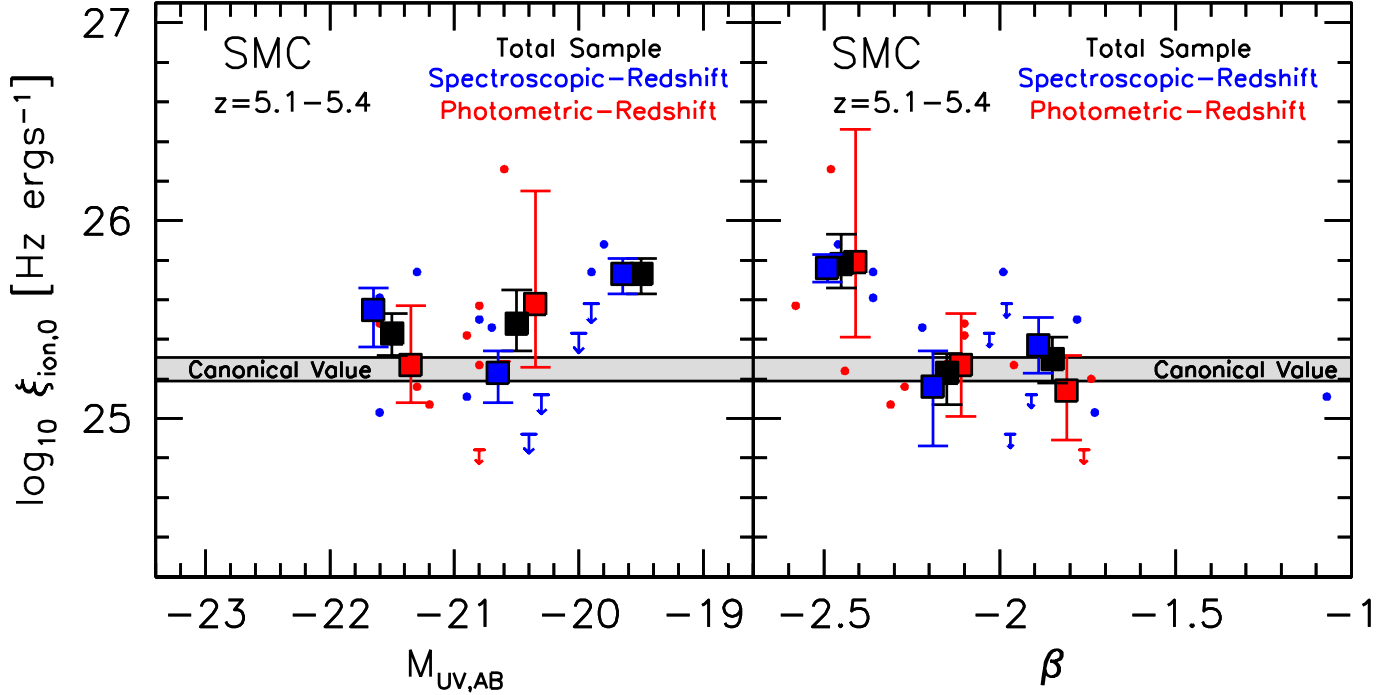}
\caption{The $\xi_{ion}$'s we have derived assuming the SMC dust law
  for $z=5.1$-5.4 galaxies from the Rasappu et al.\ (2015) selection
  shown as a function of their $UV$ luminosity and $UV$-continuum
  slope $\beta$ (\S3.2).  The blue, red, and black symbols are the
  same as on Figure~\ref{fig:xi}.  As in Figure~\ref{fig:xibeta}, we
  find that the bluest sources show particularly elevated values of
  $\xi_{ion}$ relative to canonically-assumed values.\label{fig:xi5}}
\end{figure*}

\subsection{Procedure to Derive $\xi_{ion,0}$}

The intrinsic H$\alpha$ luminosity from a galaxy is closely connected
to its total Lyman-continuum luminosity.  Based on the simulations of
Leitherer \& Heckman (1995) and assuming an escape fraction of zero
for Lyman-continuum photons into the intergalactic medium, the
H$\alpha$ luminosity $L(H\alpha)$ can be expressed in terms of the
production rate of Lyman-continuum photons $N(\textrm{H}^0)$, as
\begin{equation}
L(H\alpha) \textrm{[ergs s}^{-1}] = 1.36 \times 10^{-12} N(\textrm{H}^0)[\textrm{s}^{-1}]
\label{eq:LyCPhoton}
\end{equation}
An essentially identical conversion factor is quoted in many other
places (e.g., Kennicutt 1983, 1998; Gallagher et al.\ 1984).  The
above relationship is known to be slightly temperature and metallicity
dependent (e.g., Charlot \& Longhetti 2001).  However, in general,
these dependencies are much smaller than in converting either of these
quantities to other quantities like the star formation rate.  Overall,
the uncertainties are not expected to be larger than 15\% (0.06 dex).

It is worthwhile noting that we can make use of Eq.~\ref{eq:LyCPhoton}
even in cases where a small fraction of Lyman-continuum photons do
escape; we simply need to reinterpret $N(\textrm{H}^0)$ as referring
to those photons that do not escape from galaxies.

To make use of Eq.~\ref{eq:LyCPhoton} to derive the production rate of
Lyman-continuum photons $N(\textrm{H}^0)$ for all sources, we need to
correct the apparent H$\alpha$ fluxes we observe for the impact of
dust extinction.  For our baseline results, we derive the estimated
extinction based on the measured $UV$-continuum slopes $\beta$ for
individual sources, assuming a Calzetti et al.\ (2000) extinction law
(adopting the relation $A_{UV}=1.99(\beta+2.23)$: Meurer et
al.\ 1999), and assuming similar extinction for the nebular lines, as
for the continuum starlight.  Shivaei et al.\ (2015) demonstrated that
such a prescription produced reasonable agreement between the inferred
SFRs in the $UV$, H$\alpha$, and mid-IR (from MIPS) inferred for
galaxies at $z\sim2$.  The $\beta$'s we utilize for our dust
corrections are derived from power-law fits to the observed fluxes
(where $f_{\lambda}\propto \lambda^{\beta}$), as was first done in the
works of Bouwens et al.\ (2012b) and Castellano et al.\ (2012).  For
sources where $\beta<-2.23$ (where $A_{UV}=0$ according to the
Calzetti et al.\ (2000) dust law), we take the dust correction to be
zero.

We have made use of Eq.~\ref{eq:LyCPhoton} to derive the production
rate of Lyman-continuum photons $N(\textrm{H}^0)$ for all sources in
our samples.  We can then calculate the Lyman-continuum photon
production efficiency $\xi_{ion,0}$ as follows (with a zero subscript
to indicate that an escape fraction of zero is assumed for ionizing
photons):
\begin{equation}
\xi_{ion,0} = \frac{N(\textrm{H}^0)}{L_{UV}/f_{esc,UV}}
\end{equation}
where $L_{UV}$ is the $UV$-continuum luminosity observed for various
individual sources and $1/f_{esc,UV}$ is the dust correction to
convert the observed luminosity of a source in the $UV$-continuum to
the intrinsic luminosity (prior to the impact of dust).

\subsection{$\xi_{ion,0}$ vs. $M_{UV}$ and $\beta$}

We have presented the resultant $\xi_{ion,0}$'s for individual sources
in Figures~\ref{fig:xi} and \ref{fig:xibeta} (\textit{left panels}) as
a function of the $UV$ luminosities $M_{UV}$ of individual sources and
also the $UV$-continuum slopes $\beta$'s for sources in our
$z=3.8$-5.0 sample.  In Figure~\ref{fig:xi} and \ref{fig:xibeta}, we
present separately the results from our spectroscopic and photometric
redshift selections, as well as the results from our total sample.  In
the same figures, the mean $\xi_{ion,0}$ we have derived for sources
is also shown as a function of both $UV$ luminosity $M_{UV}$ and
$\beta$.  The same results are also presented in Table~\ref{tab:tab}.

Observational uncertainties in $\beta$ can impact the mean
$\xi_{ion,0}$ vs. $\beta$ relationship we infer through dust
corrections we apply (as we show with the thick dotted lines on
Figure~\ref{fig:xibeta}).  To determine the impact of errors in
$\beta$ on our result, we repeated our determination of $\xi_{ion,0}$
in each $\beta$ bin 300 times, but scattering the determined $\beta$'s
by a $\sigma(\beta)$ of 0.2.  [$\sigma(\beta)\sim0.2$ is the
  observational uncertainty at $z\sim4$ and $z\sim5$ for sources in
  the luminosity range we consider (Appendix B.3 of Bouwens et
  al.\ 2012b).]  None of the derived $\xi_{ion,0}$'s changed by
$>$0.05 dex as a result of adding a small scatter to $\beta$.  We
applied this small correction to the $\xi_{ion,0}$ values we report in
Figure~\ref{fig:xibeta} and Table~\ref{tab:tab}.

Given the formal size of the statistical errors on the mean
$\xi_{ion,0}$ values we derive, i.e., 0.02-0.04 dex, for different
subsamples, systematic errors likely contribute meaningfully to the
overall error budget.  Nevertheless, given the consistency of the
median H$\alpha$ equivalent width measurements derived from fitting to
the SEDs of individual sources and that derived from comparisons of
the $[3.6]-[4.5]$ colors for different redshift subsamples (see
\S3.1), systematic errors on $\xi_{ion}$ seem likely to be modest.  We
can estimate the size by comparing the median [3.6] excess derived by
Stark et al.\ (2013) using these two difference approaches, i.e., 0.37
and 0.33 mag.  The two different measures of the excess translate to
H$\alpha$ luminosities that differ at the 0.06 dex level.  We adopt
0.06 dex as our fiducial estimate of the systematic error in
$\xi_{ion}$.

For sources with redder $\beta$'s, our $\xi_{ion,0}$ results are in
good agreement with the canonical values (Table~\ref{tab:xion_lit}).
However, we derive particularly elevated $\xi_{ion,0}$'s (0.2 dex
higher than canonical assumed values) for $z=3.8$-5.0 galaxies with
the bluest $UV$-continuum slopes $\beta$ ($\beta<-2.3$).  Higher
values of $\xi_{ion}$ have indeed been predicted for those galaxies
with the bluest slopes (Bouwens et al.\ 2015b; Duncan \& Conselice
2015), so it is encouraging that our measurements provide empirical
support for these particularly elevated values of $\xi_{ion}$.  It is
worthwhile noting that this result remains the case, regardless of
what one assumes about the dust law.

Given that many early ALMA results appear to be suggesting that the
typical $z\sim5$-6 galaxy exhibits more of an SMC extinction law
(e.g., Capak et al.\ 2015) than a Calzetti et al.\ (2000) extinction
law, we also utilize the SMC dust law to correct the apparent
H$\alpha$ and $UV$-continuum fluxes and derive Lyman-continuum photon
production efficiencies $\xi_{ion,0}$ (adopting the relation
$A_{UV}=1.1\,(\beta+2.23)$ while again assuming an escape fraction of
zero for ionizing photons).\footnote{We have derived the extinction
  relation $A_{UV}=1.1(\beta+2.23)$ from the SMC observations and
  results from Prevot et al.\ (1984), Bouchet et al.\ (1985), and
  Lequeux et al.\ (1982).}  Our results are presented in the right
panels of Figures~\ref{fig:xi} and \ref{fig:xibeta}.  Interestingly
enough, the $\xi_{ion,0}$'s we derive for the SMC extinction law are
$\sim$0.07 dex higher than Calzetti and $\sim$0.1 dex higher than some
canonically assumed values (e.g., Robertson et al.\ 2015).  We discuss
comparisons with previous estimates more extensively in \S3.6.

It also makes sense for us to also derive $\xi_{ion,0}$ for the
Rasappu et al.\ (2015) $z=5.1$-5.4 samples.  We present the results in
Figure~\ref{fig:xi5} using the SMC extinction law.  Overall, the
results are in reasonable agreement with those from the Smit et
al.\ (2015b) $z=3.8$-5.0 sample.  One other striking similarity to the
$z=3.8$-5.0 results is that the bluest ($\beta<-2.3$) sources show
particularly elevated values of $\xi_{ion,0}$, again lying $\sim$0.25
dex above the canonical relationship.

Focusing on the sub-L$^*$ ($>$$-$21 mag) sources that likely play the
dominant role in reionizing the universe (e.g., Yan \& Windhorst 2004;
Bouwens et al.\ 2006, 2007, 2011; Oesch et al.\ 2010; Robertson et
al.\ 2013), we find a mean $\lxiz$ of $25.27_{-0.03}^{+0.03}$ and
$25.34_{-0.02}^{+0.02}$ for the Smit et al.\ (2015b) $z=3.8$-5.0
sample based on the Calzetti and SMC extinction laws, respectively.
For the Rasappu et al.\ (2015) $z=5.1$-5.4 sample, we find
$25.51_{-0.12}^{+0.12}$ and $25.54_{-0.12}^{+0.12}$, respectively, for
the same two extinction laws.  We emphasize that the values we derive
here do not account for a non-zero escape fraction.  We derive larger
production efficiencies (\S3.5) accounting for a positive escape
fraction.

We infer $\sim$0.3 dex intrinsic scatter in the values of $\xi_{ion}$
at a given luminosity.  In the luminosity range $-21<M_{UV,AB}<-20$,
we measure a scatter of $\sim$0.31 dex.  If one accounts for the fact
that the observational uncertainty in $\xi_{ion}$ is estimated to be
$\sim$0.18, this translates into an intrinsic scatter of $\sim$0.25
dex, very similar to the observed scatter in the main sequence of star
formation in galaxies, as inferred from H$\alpha$ (Smit et
al.\ 2015b).  See Figure~\ref{fig:spread}.  A 0.25 dex intrinsic
scatter is estimated in $\xi_{ion}$ from a simple modeling of the
fraction of sources lying above a given observed value of $\xi_{ion}$
and accounting for noise in the individual measurements of
$\xi_{ion}$.

\begin{deluxetable}{cccc}
\tablewidth{0cm}
\tabletypesize{\footnotesize}
\tablecaption{The mean $\xi_{ion,0}$'s we derive from the Inferred H$\alpha$ Flux for Galaxies of Different Luminosities and $UV$-continuum Slopes $\beta$.\label{tab:tab}}
\tablehead{
\colhead{} & \colhead{} & \multicolumn{2}{c}{$\log_{10} \bar{\xi}_{ion,0}/[\textrm{Hz~ergs}^{-1}]$\tablenotemark{a}}\\
\colhead{Subsample} & \colhead{\# Sources} & \colhead{Calzetti} & \colhead{SMC}}\\
\startdata
\multicolumn{4}{c}{$z=3.8$-5.0 Sample (Smit et al. 2015b)}\\
$-$2.6$<\beta<-$2.3 & 25 & 25.53$_{-0.07}^{+0.05}$ & 25.53$_{-0.06}^{+0.06}$\\
$-$2.3$<\beta<-$2.0 & 71 & 25.33$_{-0.04}^{+0.04}$ & 25.34$_{-0.04}^{+0.04}$\\
$-$2.0$<\beta<-$1.7 & 96 & 25.23$_{-0.05}^{+0.04}$ & 25.30$_{-0.04}^{+0.04}$\\
$-$1.7$<\beta<-$1.4 & 88 & 25.18$_{-0.04}^{+0.03}$ & 25.29$_{-0.04}^{+0.03}$\\
$-$1.4$<\beta<-$1.1 & 32 & 25.06$_{-0.05}^{+0.05}$ & 25.22$_{-0.05}^{+0.05}$\\
$-$23.0$<M_{UV}<-$22.0 & 9 & 25.08$_{-0.33}^{+0.14}$ & 25.14$_{-0.26}^{+0.14}$\\
$-$22.0$<M_{UV}<-$21.0 & 64 & 25.20$_{-0.03}^{+0.03}$ & 25.28$_{-0.03}^{+0.03}$\\
$-$21.0$<M_{UV}<-$20.0 & 195 & 25.28$_{-0.03}^{+0.03}$ & 25.34$_{-0.03}^{+0.03}$\\
$-$20.0$<M_{UV}<-$19.0 & 68 & 25.26$_{-0.06}^{+0.05}$ & 25.34$_{-0.06}^{+0.05}$\\\\
\multicolumn{4}{c}{$z=5.1$-5.4 Sample (Rasappu et al. 2015)}\\
$-$2.6$<\beta<-$2.3 & 7 & --- & 25.78$_{-0.12}^{+0.15}$\\
$-$2.3$<\beta<-$2.0 & 6 & --- & 25.23$_{-0.16}^{+0.10}$\\
$-$2.0$<\beta<-$1.7 & 9 & --- & 25.30$_{-0.11}^{+0.12}$\\
$-$22.0$<M_{UV}<-$21.0 & 6 & --- & 25.43$_{-0.11}^{+0.10}$\\
$-$21.0$<M_{UV}<-$20.0 & 13 & --- & 25.48$_{-0.14}^{+0.17}$\\
$-$20.0$<M_{UV}<-$19.0 & 3 & --- & 25.73$_{-0.10}^{+0.08}$\\
\enddata
\tablenotetext{a}{Assumes that the escape fraction is zero.  The
  estimated $\xi_{ion,0}$'s would be $\sim$0.03 dex higher if we
  account for a positive escape fraction and suppose that galaxies
  dominate the observed ionizing emissivity at $z\sim4$-5.  See \S3.5.
  In addition to the formal uncertainties quoted on $\xi_{ion}$, the
  derived values are likely subject to a small systematic error, i.e., 0.06 dex
  (see \S3.3).}
\end{deluxetable}

\begin{figure}
\epsscale{1.1}
\plotone{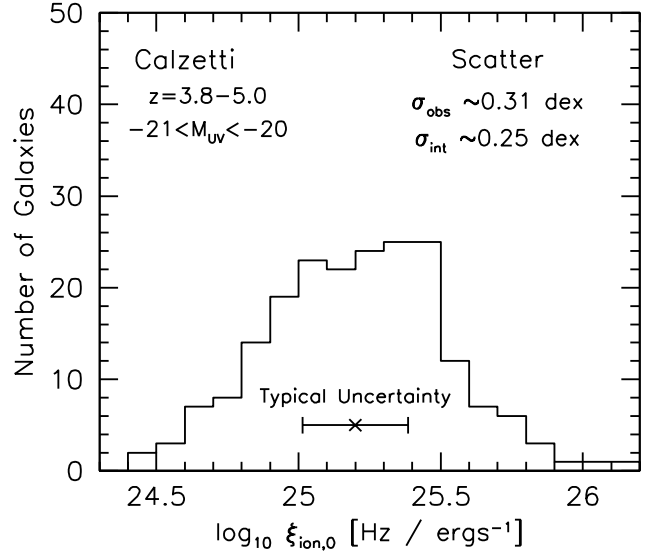}
\caption{Distribution of $\xi_{ion}$'s estimated from the observations
  for $z=3.8$-5.0 galaxies in the luminosity range $-21<M_{UV,AB}<-20$
  assuming a Calzetti dust law.  For 10\% of the sources where
  H$\alpha$ is not detected at $1\sigma$ significance, the individual
  $\xi_{ion}$ values are presented at their $1\sigma$ upper limits on
  the histogram.  The observed scatter in this distribution is
  $\sim$0.31 dex.  Given that the typical uncertainty in individual
  estimates of $\xi_{ion}$ is $\sim$0.18 dex (shown as a horizontal
  error bar with respect to the median $\xi_{ion}$ plotted as a
  cross), this implies an intrinsic scatter of $\sim$0.25 dex, very
  similar to the scatter around the main sequence of star formation in
  galaxies, as estimated by Smit et al.\ (2015b) based on the inferred
  H$\alpha$ fluxes.  Essentially an identical intrinsic scatter is
  derived modeling the cumulative distribution of $\xi_{ion}$ values
  accounting for individual observational errors.  See
  \S3.3.\label{fig:spread}}
\end{figure}

\subsection{Dust Extinction Impacting the Nebular vs. Stellar Continuum Light}

In addition to uncertainties that directly regard the dust law, it is
also unclear whether emission lines suffer more extinction than
stellar continuum light due to a significant dust mass in nebular
regions of galaxies.  While the nebular continuum is known to be more
extincted than the stellar continuum in the local universe, i.e.,
$A_{V,stellar} = 0.44 A_{V,gas}$ (Calzetti et al.\ 1997, 2000), select
results at $z\sim2$ suggests that this is not true for all $z\sim2$
galaxies and many exhibit $A_{V,stellar} = A_{V,gas}$ (e.g., Erb et
al.\ 2006; Reddy et al.\ 2010, 2015; but see also F{\"o}rster
Schreiber et al. 2009; Kashino et al.\ 2013; Price et al.\ 2014).

We rederived $\xi_{ion,0}$ for the individual sources in our samples
assuming that nebular lines suffer a 2.3$\times$ higher dust
obscuration.  In this case, the derived $\xi_{ion,0}$ would be 0.09
dex and 0.02 dex higher for the Calzetti and SMC dust laws,
respectively.  We do not correct our baseline determinations for this
effect given evidence from other studies (e.g., Shivaei et al.\ 2015)
that such a correction is not clearly necessary for achieving
agreement between $UV$, H$\alpha$, and mid-IR-based SFR estimates.

\subsection{Sensitivity to the Assumed Escape Fraction}

A separate factor which impacts the Lyman-continuum photon production
efficiency $\xi_{ion}$ is the escape fraction of ionizing photons we
assume.  If the escape fraction is larger than zero, then some
fraction of the ionizing photons are escaping from a galaxy without
having an impact on the number of ionized hydrogen atoms within a
galaxy and also on its H$\alpha$ luminosity.  The implication is that
those photons which do not escape must be even more rich in
Lyman-continuum photons (per unit $UV$ luminosity) than we would infer
if no radiation at all was escaping.

Following the work of Kuhlen \& Faucher-Gigu{\`e}re (2012), we can set
upper limits on the escape fraction of ionizing radiation at
$z\sim4.4$ from galaxies by comparing the $UV$ luminosity density
integrated to various limiting luminosities with measurements of the
ionizing emissivity $\dot{N}_{ion}$.  The relevant equation is
\begin{equation}
\dot{N}_{ion} = f_{esc} \xi_{ion} \rho_{UV}
\end{equation}
(e.g., Robertson et al.\ 2013; see also Kuhlen \& Faucher-Gigu{\'e}re
2012).  The ionizing emissivity has been measured at $z\sim4.4$ based
on observations of the Lyman-$\alpha$ forest which constrain both the
ionizing background and the mean-free path of ionizing photons;
interpolating between the $z\sim4$ and $z\sim4.75$ measurements of
Becker \& Bolton (2013), we adopt a value of $10^{50.92\pm0.45}$
s$^{-1}$ Mpc$^{-3}$.  If we assume that the $UV$ LF has a faint-end
cut-off at $-13$ mag, then the integrated luminosity we estimate by
interpolating between the $z\sim3.8$ and $z\sim4.9$ LF results from
Bouwens et al.\ (2015a) is $10^{26.56\pm0.06}$ ergs s$^{-1}$Hz$^{-1}$
Mpc$^{-3}$.  $\xi_{ion}$ represents the Lyman-continuum photon
production efficiency in the presence of a non-zero escape fraction
and is equal to $\xi_{ion,0}/(1-f_{esc,LyC})$.  Meanwhile, $f_{esc}$
represents the so-called relative escape fraction $f_{esc} =
f_{esc,LyC}/f_{esc,UV}$ where $f_{esc,LyC}$ and $f_{esc,UV}$ represent
the escape fraction at Lyman-continuum and $UV$-continuum wavelengths,
respectively (see e.g. Steidel et al.\ 2001; Shapley et al.\ 2006;
Siana et al.\ 2010).  The expanded expression is $f_{esc,LyC}
\xi_{ion,0} / (1-f_{esc,LyC}) / f_{esc,UV}$ =
$\dot{N}_{ion}/\rho_{UV}$ if galaxies provide the dominant
contribution to the ionizing emissivity at $z=4$-5.

If we consider the case that $\beta\sim -2$ (which is typical for
sources in the magnitude range we consider: Bouwens et al.\ 2014), the
escape fraction of $UV$-continuum photons $f_{esc,UV}$ is 0.8 adopting
the SMC dust extinction law (where $\lxiz = 25.34$).  This translates
to $f_{esc,LyC}$ being equal to $0.08_{-0.05}^{+0.12}$.  The quoted
errors here allow for the full range of systematic errors permitted in
the ionizing emissivity results of Becker \& Bolton (2013).  If the
dust curve is Calzetti and $f_{esc,UV}=0.7$ for a $\beta\sim-2$
source, then $f_{esc,LyC}=0.08_{-0.05}^{+0.12}$.

These estimated escape fractions imply that $\xi_{ion}$ can be at most
$\log_{10} 1/(1-0.08_{-0.05}^{+0.12})$ $\sim$ 0.03$_{-0.02}^{+0.06}$
dex larger than what we measure for $\xi_{ion,0}$ from the inferred
H$\alpha$ fluxes.

\begin{deluxetable}{cc}
\tablewidth{0cm} \tabletypesize{\footnotesize} \tablecaption{Current
  measurements of $\xi_{ion,0}$ vs. those previously assumed in
  reionization models.\label{tab:xion_lit}}
\tablehead{\colhead{Empirical Determination} & \colhead{$\log_{10} \xi_{ion,0}$}
  \\ \colhead{} & \colhead{[Hz$~$ergs$^{-1}$]}}
\startdata 
\multicolumn{2}{c}{Current Determinations}\\
\multicolumn{2}{c}{$z=3.8$-5.0}\\ 
Fiducial Determination (SMC Dust)\tablenotemark{a,$\dagger$} & 25.34$_{-0.02}^{+0.02}$\tablenotemark{b,$\dagger$} \\ 
Calzetti Dust Extinction\tablenotemark{a,$\dagger$} & 25.27$_{-0.03}^{+0.03}$\tablenotemark{b,$\dagger$} \\ \\
\multicolumn{2}{c}{$z=5.1$-5.4}\\ 
Fiducial Determination (SMC Dust)\tablenotemark{a,$\dagger$} & 25.54$_{-0.12}^{+0.12}$\tablenotemark{b,$\dagger$} \\ 
Calzetti Dust Extinction\tablenotemark{a,$\dagger$} & 25.51$_{-0.12}^{+0.12}$\tablenotemark{b,$\dagger$} \\ \\
\multicolumn{2}{c}{Previous Estimates}\\ 
$z=7.045$: Stark et al.\ (2015) & $25.68_{-0.19}^{+0.27}$\tablenotemark{c}\\\\
\multicolumn{2}{c}{Based on Previously Inferred $L_{H\alpha}$ and SFR$_{UV}$ Values}\\ 
$z\sim4.5$: Shim et al.\ (2011) & 25.72$_{-0.04}^{+0.04}$\tablenotemark{e,$\dagger$}\\
$z\sim4.5$: Marmol-Queralto et al.\ (2016) & 25.08$_{-0.04}^{+0.04}$\tablenotemark{f,$\dagger$}\\\\
\multicolumn{2}{c}{Based on Canonical Conversion Factors}\\
Kennicutt (1998)\tablenotemark{d} & 25.11\\\\
\multicolumn{2}{c}{Previously Suggested Values}\\ 
Madau et al.\ (1999) & 25.3\\
Robertson et al.\ (2013) & 25.20 \\ 
Robertson et al.\ (2015) & 25.24 \\ 
Topping \& Shull (2015) & 25.4$\pm$0.2\tablenotemark{g}\\ 
Bouwens et al.\ (2015b,c) & 25.46 \\ 
Kuhlen \& Faucher-Gigu{\`e}re (2012) & 25.30 \\
 & 25.00 - 25.60 \\
Bouwens et al.\ (2012a) & 25.30 \\ 
Finkelstein et al.\ (2012b) & 25.28\tablenotemark{h} \\ 
Duncan \& Conselice (2015) Model A & 25.18
\enddata
\tablenotetext{$\dagger$}{In addition to the formal uncertainties quoted on $\xi_{ion}$, the
  derived values are likely subject to a small systematic error, i.e., 0.06 dex
  (see \S3.3).}
\tablenotetext{a}{$E(B-V)_{neb}=E(B-V)_{stellar}$}
\tablenotetext{b}{If we assume that galaxies provide the dominant
  contribution to the cosmic ionizing emissivity at $z>4$, we 
  require a non-zero Lyman-continuum escape fraction from galaxies.  
  If we account for this, the $\xi_{ion}$'s we derive would be 0.03 dex 
  higher (\S3.5).}
\tablenotetext{c}{Constraints on $\xi_{ion}$ using the
  detected flux in the CIV$\lambda$1548 emission line.}
\tablenotetext{d}{Implied value of $\xi_{ion}$ using the conversion
  factors Kennicutt (1998) quote for converting $UV$ and
  H$\alpha$ luminosities into star formation rates.}
\tablenotetext{e}{Implied value of $\xi_{ion}$ based on the median UV
  to $H\alpha$ SFRs quoted by Shim et al.\ (2011).  As Shim et
  al.\ (2011) consider those sources with significant evidence for
  H$\alpha$ emission, their $\xi_{ion}$ might be expected to be 
  significantly higher than what we derive.}
\tablenotetext{f}{Implied value of $\xi_{ion}$ based on the median UV
  to $H\alpha$ SFRs quoted by Marmol-Queralto et al.\ (2016).  Using
  our own results, we estimate this value to be lower than our own
  fiducial determination using a sub-$L^*$ sample, since this value is 
  the median rather than the mean (impact of 0.08 dex) and is derived using 
  a high-mass ($>$10$^{9.5}$ $M_{\odot}$) subsample (impact of 0.08 dex).}
\tablenotetext{g}{Converted using Salpeter IMF}
\tablenotetext{h}{At face value, the 13\% factor advocated here is
  similar to values suggested in previous work (Finkelstein et
  al.\ 2012b), but the correspondence is accidental given significant
  changes in the preferred values for both $\dot{N}_{ion} (z=6)$ and
  $\xi_{ion}$ (as well as $\rho_{UV}$) over the last three years.  Of
  particular note, Bouwens et al.\ (2015b) have presented evidence
  based on a simple modeling of the ionizing emissivity evolution that
  $\dot{N}_{ion}(z=6)$ is likely $\sim$0.3-0.5 dex higher than
  concluded by Bolton \& Haehnelt (2007) using more direct methods.}
\end{deluxetable}

\subsection{Comparison with Previous Estimates}

The literature is full of a wide variety of observational,
theoretical, and empirical results for the Lyman-continuum photon
production efficiency $\xi_{ion}$.  Table~\ref{tab:xion_lit} provides
a useful summary of many of them.

\subsubsection{Suggested Values from Stellar Population Models}

Many of the first estimates were based on the results of standard
stellar population models (e.g., Bruzual \& Charlot 2003) at normal or
slightly sub-solar metallicities (Madau et al.\ 1999; Bouwens et
al.\ 2012a; Kuhlen \& Faucher-Gigu{\`e}re 2012; Finkelstein et
al.\ 2012b).  Many relevant models (Schaerer 2003: see also Bruzual \&
Charlot 2003) suggested $\lxi$ values of 25.20 at solar metallicites.

Use of the conversion factors from Kennicutt (1998) indicate
25.11 for the value of $\lxi$.

\subsubsection{Inferred from the Measured $UV$-continuum Slopes}

$\xi_{ion}$ has also been estimated based on the mean $UV$-continuum
slopes $\beta$ derived in a number of different observational studies
(Robertson et al.\ 2013, 2015; Bouwens et al.\ 2015b; Duncan \&
Conselice 2015).  Robertson et al.\ (2013) attempted to match
$\beta\sim-2$ measurements by Dunlop et al.\ (2013) and estimated
$\lxi$ to be 25.20.

Meanwhile, Bouwens et al.\ (2015b) found $\lxi$ to be 25.46 using a
similar procedure to Robertson et al.\ (2013) but with aim of matching
the mean $\beta$ value of $\sim-2.3$ derived by Bouwens et al.\ (2014)
for fainter $z\sim7$ galaxies.  Duncan \& Conselice (2015) also made
note of the bluer $\beta$ values derived for fainter galaxies by
Bouwens et al.\ (2012b), Bouwens et al.\ (2014), Rogers et
al.\ (2014), and Finkelstein et al.\ (2012a) and also the bluer
$\beta$'s derived for galaxies at higher redshifts (Bouwens et
al.\ 2012b, 2014; Finkelstein et al.\ 2012a; Hathi et al.\ 2013;
Kurczynski et al.\ 2014: see also Wilkins et al.\ 2015).

\subsubsection{From near-UV Spectroscopy}

Another recent estimate of the Lyman-continuum photon production
efficiency $\lxi$ is $\lxi=25.68_{-0.19}^{+0.27}$ and came from a
recent analysis by Stark et al.\ (2015) of a lensed $z=7.045$ galaxy
A1703-zD6 behind Abell 1703 (Bradley et al.\ 2012).  Stark et
al.\ (2015) obtained this result from an analysis of the near-infrared
spectrum they collected of this source, which excitedly enough shows
the detection of a prominent CIV$\lambda$1548 emission line.  Stark et
al.\ (2015) found they could not reproduce the observed properties of
this line, as well as the other properties of the source, without a
high production efficiency of Lyman-continuum photons, i.e.,
$\lxi=25.68_{-0.19}^{+0.27}$ and also including photons with energies
$>$20-30 eV.

While higher than the mean value we obtain for our sample, the
$\log_{10} \xi_{ion}$ Stark et al.\ (2015) derive for this source is
actually quite consistent with what we measure for the bluest sources
in our selection 25.53$_{-0.06}^{+0.06}$ (Table~\ref{tab:tab}),
especially considering the intrinsic variation in $\xi_{ion}$ that
appears to be present across individual sources at $z\sim4$-5
(Figure~\ref{fig:spread}).  Future spectroscopy on A1703-zD6 probing
H$\alpha$ and H$\beta$ emission with NIRSPEC and MIRI on JWST should
provide for an independent test of the $\lxi$ measured by Stark et
al.\ (2015).

\subsubsection{Based on Previous $L_{H\alpha}$ Measurements}

Previous work even provide implicit determinations of the
Lyman-continuum photon production efficiency $\xi_{ion}$ for
$z\sim4$-5 galaxies based on the inferred H$\alpha$ and $UV$
luminosities, even though it was specifically represented in these
terms.

One such study is from Shim et al.\ (2011).  Shim et al.\ (2011)
provide H$\alpha$ luminosities $L_{H\alpha}$ and $UV$-based SFRs for
74 sources over the GOODS-North and GOODS-South fields with
spectroscopic redshifts in the range $z=3.8$-5.0.  We can compute the
equivalent $\xi_{ion}$ from Eq. (2) using their quoted values for
$L_{H\alpha}$, SFR$_{UV}$, and $\beta$ and converting SFR$_{UV}$ into
a UV luminosity using the relations tabulated by Kennicutt (1998).  We
estimate a value of 25.64$_{-0.04}^{+0.01}$ and
25.72$_{-0.04}^{+0.04}$ for the median and mean value of $\lxi$,
respectively, assuming an SMC dust law.  We would expect the Shim et
al.\ (2011) values to be higher than our values, since they only
considered sources which showed clear evidence for an H$\alpha$
emission line in the photometry.  The use of such a selection would
bias their measured $\xi_{ion}$ towards higher values.

Alternatively, if we make use of the median inferred SFRs from
Marmol-Queralto et al.\ (2016), we find $\lxi$ to be equal to 25.08,
which is lower than what we derive by 0.19 dex.  There appear to be
two reasons for this difference.  First, Marmol-Queralto et
al.\ (2016) quote the median value whereas we quote the mean.  Second,
Marmol-Queralto et al.\ (2016) consider a higher-mass subsample, i.e.,
$>10^{9.5}$ $M_{\odot}$, than what we consider.  If we use our own
sample as a guide, these two choices would lower our derived value of
$\lxi$ by 0.08 dex and 0.08 dex, respectively, which when summed
essentially match the difference between the two results.  See also
brief discussion in Smit et al.\ (2016) regarding the reasonable
overall agreement with the Marmol-Queralto et al.\ (2016)
determinations of the H$\alpha$ EWs.

\begin{figure}
\epsscale{1.1}
\plotone{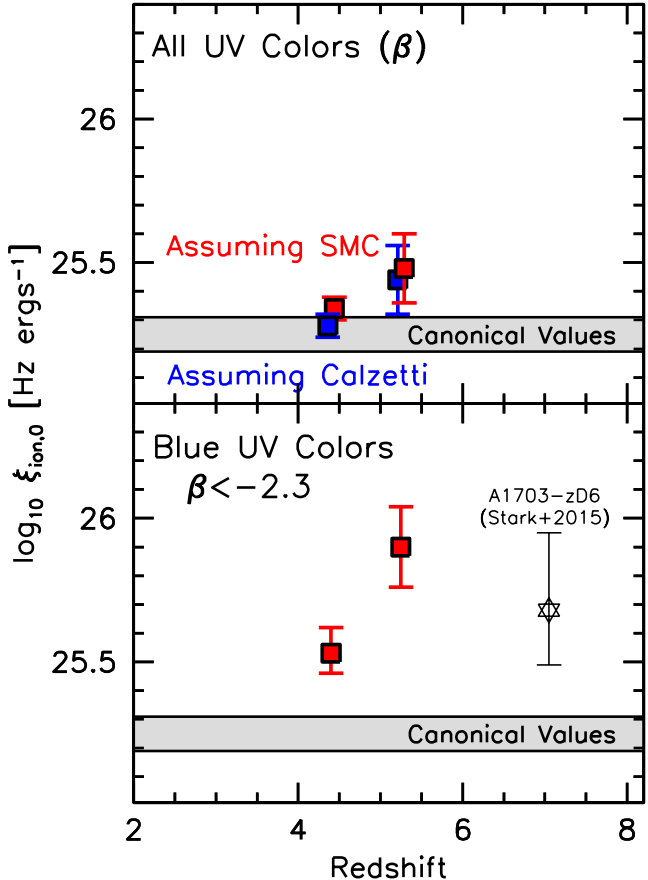}
\caption{(\textit{upper}) Mean Lyman-continuum photon production
  efficiency $\xi_{ion}$ estimated for star-forming galaxies at
  $z\sim4.4$ and $z\sim5.25$ from the inferred H$\alpha$ flux using
  both the Calzetti et al.\ (2000) and SMC dust laws.  The plotted
  values are not corrected for escaping Lyman-continuum photons (if
  this fraction is significant).  Also shown are the canonical
  $\xi_{ion}$ values utilized in the literature to model the impact of
  galaxies on the reionization of the universe.  Our derived values
  for $\xi_{ion}$ are either consistent or $1\sigma$ higher than
  canonically assumed values.  (\textit{lower}) Mean Lyman-continuum
  photon production efficiency $\xi_{ion}$ estimated for the bluest
  ($\beta<-2.3$) star-forming galaxies at $z\sim4.4$ and $z\sim5.25$
  from the inferred H$\alpha$ flux.  Also shown is the $\xi_{ion}$
  derived by Stark et al.\ (2015) for one blue $\beta=-2.4$ $z=7.045$
  galaxy from the observed CIV$\lambda$1548 line.  $\xi_{ion}$ is
  inferred to be consistently higher (by $\sim$0.2-0.3 dex) than has
  been canonically-assumed for the star-formation population as a
  whole at $z>6$ in reionization modeling.\label{fig:evolution}}
\end{figure}

\begin{figure*}
\epsscale{1.1}
\plottwo{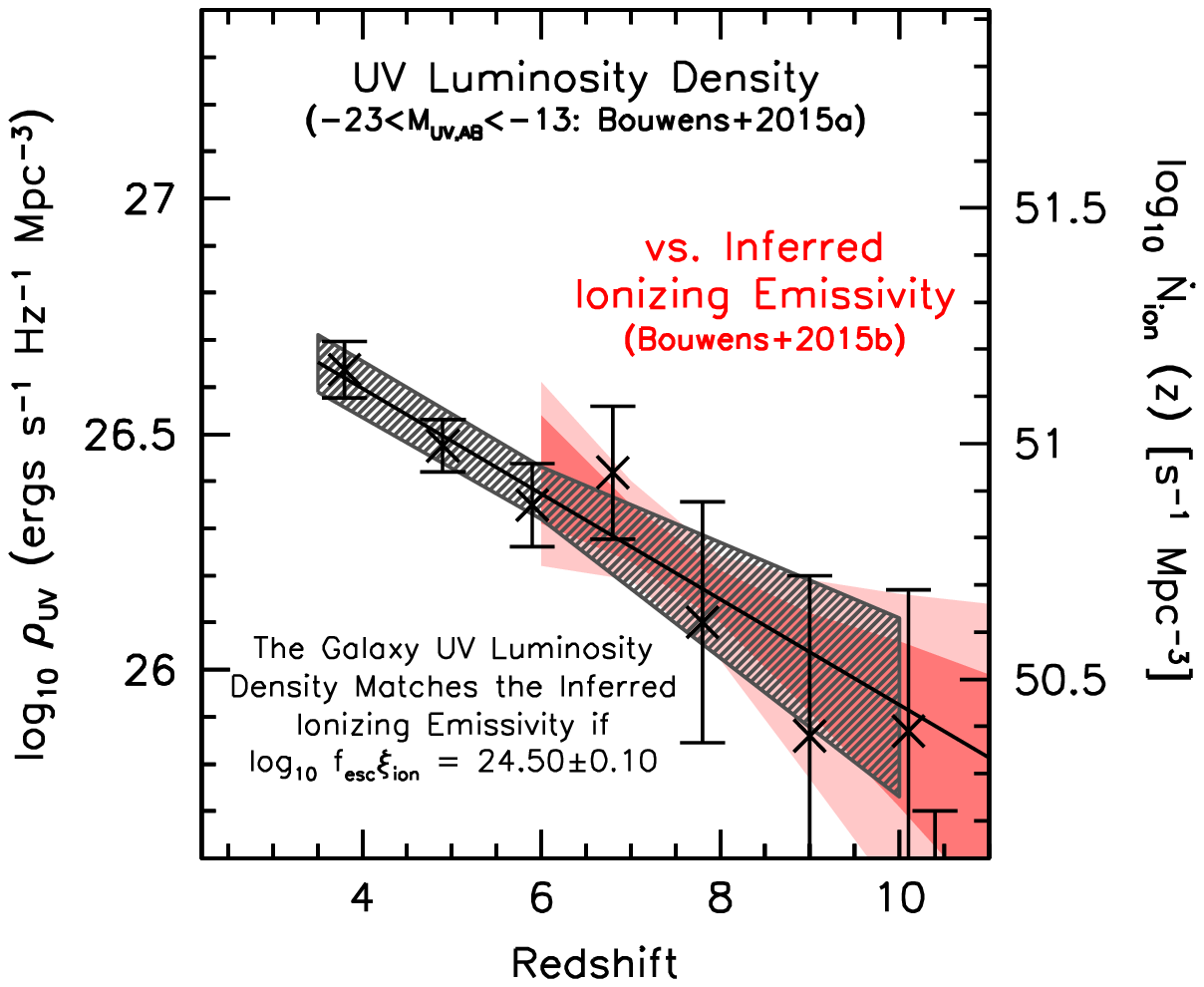}{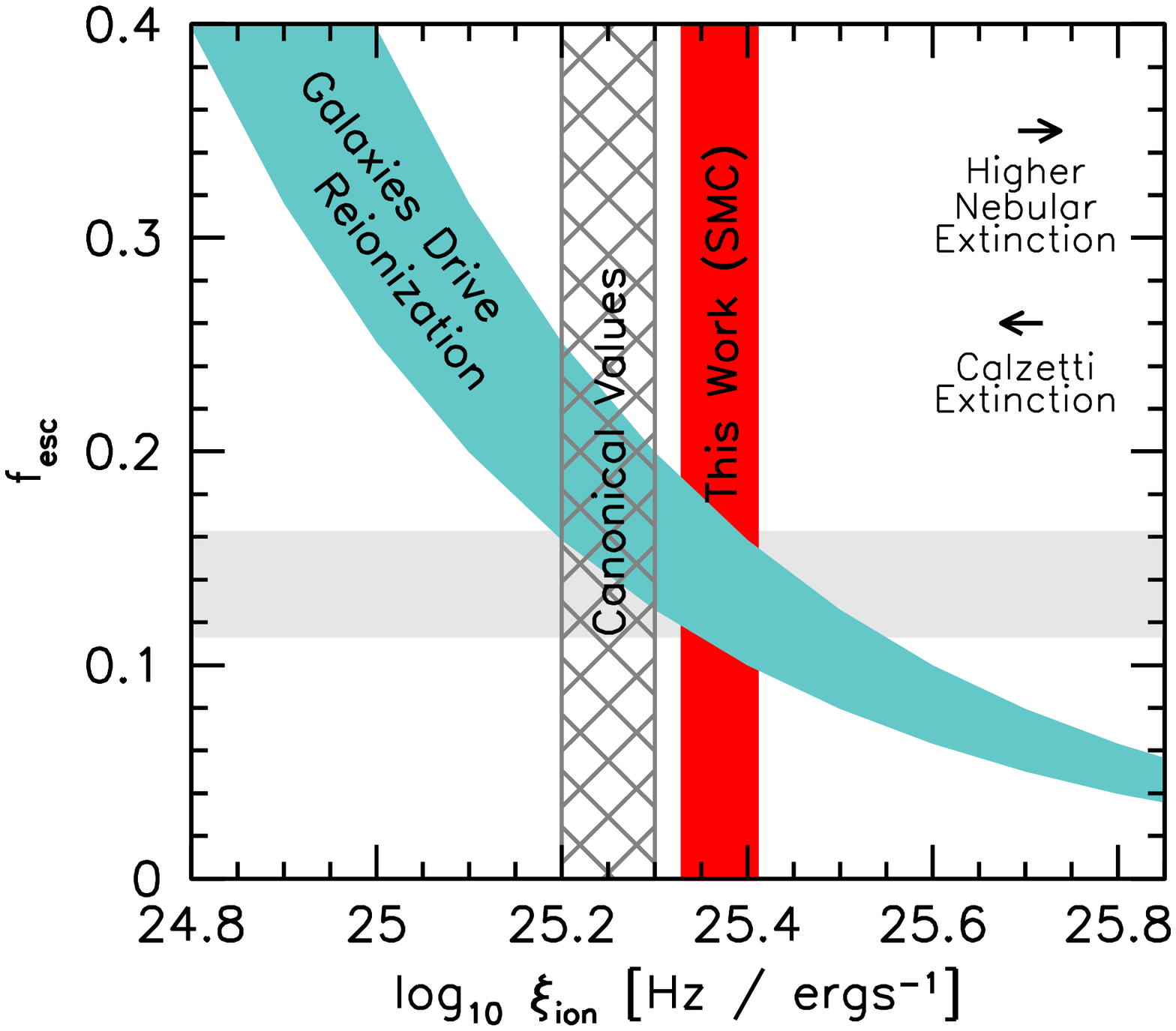}
\caption{Implications from our current $\xi_{ion}$ results for the
  escape fraction $f_{esc}$ in $z>6$ galaxies (assuming similar
  $\xi_{ion}$'s in $z>6$ galaxies as at $z=4$-5).  (\textit{left})
  Determinations of the $UV$ luminosity densities at $z=4$-10
  integrated to $-13$ mag (\textit{black crosses with error bars and
    the shaded regions giving parameterizing fit results}: Bouwens et
  al.\ 2015b) from several recent LF determinations (Bouwens et
  al.\ 2015a; Ishigaki et al.\ 2015; Oesch et al.\ 2015) compared to
  the inferred evolution (\textit{dark red and light red contours give
    the 68\% and 95\% confidence intervals}) of the cosmic ionizing
  emissivity from $z=6$-12 (Bouwens et al.\ 2015b: see also Mitra et
  al.\ 2015).  As demonstrated first by Robertson et al.\ (2013, 2015)
  and later by Bouwens et al.\ (2015), $\lxif=24.50$ if galaxies drive
  the reionization of the universe, the faint-end cut-off to the LF is
  $-13$, and the clumping factor $C$ is 3; higher values of
  $\xi_{ion}$ directly translate into lower required values for
  $f_{esc}$.  (\textit{right}) Implied constraints on the relative
  escape fraction $f_{esc} = f_{esc,LyC}/f_{esc,UV}$ for $z\geq6$
  galaxies (\textit{horizontal light gray region}) that we can set on
  the basis of our measured $\xi_{ion}$ adopting the relationship
  $\lxif=24.50\pm0.10$ (\textit{shaded cyan region}: \S3.5).  The dark
  red regions give the measured $\xi_{ion}$ values our analysis
  prefers at 68\% confidence adopting a SMC extinction law and
  assuming that galaxies drive the reionization of the universe.  If
  we assume that the dust extinction follows the Calzetti dust law,
  our measured $\xi_{ion}$ would be 0.07 dex lower; however, if we
  assume that line emission from nebular regions suffer from more dust
  extinction than stellar continuum light, our measured $\xi_{ion}$
  would be 0.02-0.09 dex higher.  The vertical hatched grey region
  indicates the Lyman-continuum photon production efficiencies
  $\xi_{ion}$ assumed in typical models
  (Table~\ref{tab:xion_lit}).  \label{fig:fesc}}
\end{figure*}

\begin{deluxetable}{cccc}
\tablewidth{0cm}
\tabletypesize{\footnotesize}
\tablecaption{Required Values of $f_{esc}$ for different $M_{lim}$ and clumping factors $C_{HII}$ assuming that galaxies drive the reionization of the universe.\tablenotemark{a}\label{tab:fesc}}
\tablehead{
\colhead{} & \multicolumn{3}{c}{Required $f_{esc}$ ($=f_{esc,LyC}/f_{esc,UV}$)} \\
\colhead{} & \multicolumn{3}{c}{$\xi_{ion}= 10^{25.37}$\tablenotemark{$\ddagger$} $\textrm{Hz~ergs}^{-1}$} \\
\colhead{$C_{HII}$} & \colhead{$M_{lim}=-17$} & \colhead{$M_{lim}=-13$} & \colhead{$M_{lim}=-10$}}
\startdata
2.0 & 0.31$_{-0.06}^{+0.08}$ & 0.12$_{-0.02}^{+0.03}$ & 0.08$_{-0.02}^{+0.02}$\\
3.0 & 0.35$_{-0.07}^{+0.09}$ & 0.13$_{-0.03}^{+0.03}$ & 0.09$_{-0.02}^{+0.02}$\\
5.0 & 0.41$_{-0.09}^{+0.11}$ & 0.16$_{-0.03}^{+0.04}$ & 0.10$_{-0.02}^{+0.03}$\\
2.4\tablenotemark{$\dagger$} & 0.33$_{-0.07}^{+0.09}$ & 0.13$_{-0.03}^{+0.03}$ & 0.08$_{-0.02}^{+0.02}$
\enddata
\tablenotetext{$\dagger$}{Redshift Dependence found in the hydrodynamical simulations of Pawlik et al.\ (2009).}
\tablenotetext{$\ddagger$}{$10^{25.37}$ Hz$~$ergs$^{-1}$ is the approximate Lyman-continuum photon production efficiency $\xi_{ion}$, if the dust curve is SMC and after accounting for the fact that the $\xi_{ion}$'s estimated from the inferred H$\alpha$ fluxes do not account for the $\sim$9-10\% of the Lyman-continuum photons that escape from galaxies (assuming that galaxies dominate the ionizing emissivity at $z>4$).  See \S3.5.}
\tablenotetext{a}{These $f_{esc}$ factors can be derived from Eq.~(\ref{eq:convf}).  Importantly, we can also quote uncertainties on the estimated $f_{esc}$'s, which follow from the $1\sigma$ error estimate ($\sim$0.1 dex) on the conversion factor $10^{24.50}$ Hz$~$ergs$^{-1}$ from $UV$ luminosity density $\rho_{UV}$ to the equivalent ionizing emissivity $\dot{N}_{\rm ion}$ (Bouwens et al.\ 2015b).}
\end{deluxetable}

\subsection{Redshift Evolution}

It is worthwhile investing the apparent evolution of $\xi_{ion}$ with
cosmic time.  This is shown in Figure~\ref{fig:evolution} for both our
mean $\xi_{ion}$ derived from our samples as a whole (\textit{upper
  panel}) and making exclusive use of sources with the bluest measured
$\beta$'s, i.e., $\beta<-2.3$ (\textit{lower panel}).  Results are
presented in assuming both Calzetti and SMC extinction laws.  The
results are consistent with (or perhaps $1\sigma$ higher than) what
has been canonically assumed for $\xi_{ion}$ in standard reionization
models (e.g., Kuhlen \& Faucher-Gigu{\`e}re 2012; Robertson et
al.\ 2013).

Interestingly enough, the $\xi_{ion}$'s we estimate for the bluest
subsample of galaxies are consistently higher than canonically-assumed
values, but are consistent with what Stark et al.\ (2015) estimate for
one blue $\beta\sim-2.4$ source at $z=7.045$.  This suggests that
those galaxies with the bluest $UV$ colors may be consistently the
most efficient at producing the Lyman-continuum photons capable of
reionizing the universe.

\section{Discussion}

In the present work, we have used new measurements of the H$\alpha$
luminosities in $z=3.8$-5.4 galaxies to estimate the Lyman-continuum
photon production efficiency $\xi_{ion}$.  Assuming that early results
with ALMA (e.g., Capak et al.\ 2015) at $z=5$-6 are correct and dust
emission is more SMC like, we derive a Lyman-continuum photon
production efficiency $\lxiz$ of $25.34_{-0.02}^{+0.02}$ at
$z=3.8$-5.0.  Higher values (by $\sim$0.03-dex) would be expected if
the escape fraction is non-zero and galaxies contribute meaningfully
to the observed ionizing emissivity.

Our results for $\xi_{ion}$ are consistent with standard assumptions
in canonical models.  Nevertheless, for the SMC dust law preferred by
early ALMA result, they are suggestive of even higher ($\sim$0.1 dex)
values for $\xi_{ion}$ than traditionally assumed.  If the $\xi_{ion}$
values are indeed higher than in canonical modeling, it could have a
number of important implications.  It would impact our understanding
of (1) the stellar populations in $z>2$ galaxies, (2) the
required/allowed escape fraction in high-redshift galaxies, and (3)
the methodology for constraining the escape fraction in the future
JWST mission.

\subsection{Implications for the Stellar Populations of $z>2$ Galaxies}

The present results show that $z>3$ galaxies produce Lyman-continuum
photons at the same rate as (or higher than) expected in conventional
stellar population models.  In the case that $\xi_{ion}$ is higher
than conventional models, we could try to explain this result by
adopting particularly bursty star-formation histories for $z>2$
galaxies.

Such bursty star formation histories are disfavored by several recent
results.  Specifically, Oesch et al.\ (2013) find that the $J-[4.5]$
color distribution (providing a measure of the Balmer-break amplitude)
shows a generally normal-looking distribution, with a peak at 0.4 mag,
which is exactly where one would expect the peak to lie using
semi-analytic models based on the Millenium simulations.  Secondly,
Smit et al.\ (2015b) find a strong correlation between $UV$ and
$H\alpha$-based specific star formation rates, pointing towards a
generally monotonic growth in the SFR and limited variations in the
SFR on $\sim$10-20 Myr time scales.

A more credible explanation for a high production efficiency for
Lyman-continuum photons (if confirmed to be the case with smaller
uncertainties) would involve evolution in the IMF of galaxies or
evolution in the way that high-mass stars evolve at early times.
There have been several suggestions that such changes are indeed found
in the newer generations of stellar evolution models.  It has become
clear that massive stars are predominantly found in binaries (Sana et
al.\ 2012) and rotate with a wide range of rotation rates (e.g.,
Ramirez-Agudelo et al.\ 2013).  The new models that account for these
effects predict a higher production efficiency for Lyman-continuum
photons at early times when the average metallicity was lower (Yoon et
al.\ 2006; Eldridge \& Stanway 2009, 2012; Levesque et al. 2012; de
Mink et al.\ 2013; Kewley et al. 2013; Leitherer et al. 2014; Sz{\'e}csi
et al.\ 2015; Gr{\"a}fener et al.\ 2015; Stanway et al. 2015).

\subsection{Implications for the Escape Fraction}

The $\xi_{ion}$'s we derive from the observations are slightly larger
than preferred in some previous work on reionization, particularly in
the case of the SMC extinction law, and so it is useful for us to
consider the impact this may have on the allowed escape fraction for
$z>6$ galaxies, assuming similar $\xi_{ion}$'s in reionization-era
galaxies.

As demonstrated in previous work (e.g., Robertson et al.\ 2013), we
can set strong constraints on the escape fraction $f_{esc}$ if we know
$\xi_{ion}$.  [This assumes fiducial choices for other variables, i.e,
  a clumping factor $C=<(n_H)^2>/<n_H>^2)$ of 3 and fiducial faint-end
  cut-off to the LF of $-13$.]  The implicit constraint in Robertson
et al.\ (2013, 2015) is for $\lxif$ to equal 24.50.  Bouwens et
al.\ (2015) found essentially identical constraints on $f_{esc} \xi_{ion}$ in a follow-up analysis, but presented this constraint in a
generalized form to a wider range of faint-end cut-offs to the $UV$ LF
$M_{lim}$ and clumping factors $C$:
\begin{equation}
f_{esc}\xi_{\rm ion} f_{corr}(M_{lim}) (C/3)^{-0.3} = ~~~~~~~~~~~~~~~~~~~~
\label{eq:convf}
\end{equation}
\begin{displaymath}
~~~~~~~~~~~~~~~~~~10^{24.50\pm0.10} \textrm{Hz}\,\textrm{ergs}^{-1}
\end{displaymath}
where $f_{corr} (M_{lim}) =
10^{0.02+0.078(M_{lim}+13)-0.0088(M_{lim}+13)^2}$ corrects
$\rho_{UV}(z=8)$ derived to a faint-end limit of $M_{lim}=-13$ mag to
account for different $M_{lim}$'s (\textit{left panel} of
Figure~\ref{fig:fesc}).

If we adopt a faint-end cut-off to the $UV$ LF of $-$13 mag, take the
clumping factor $C$ to be 3 (as favored by Pawlik et al.\ 2009: see
also Bolton \& Haehnelt 2007, Shull et al.\ 2012, Finlator et
al.\ 2012; Pawlik et al.\ 2015) and alternatively take $\lxi$ to be
$25.37_{-0.03}^{+0.02}$ and $25.31_{-0.03}^{+0.03}$ as appropriate for
SMC and Calzetti extinction, we estimate $f_{esc}$ to be 0.13 and
0.14, respectively (\textit{right panel} of Figure~\ref{fig:fesc}).
Equivalent results are also presented in Table~\ref{tab:fesc} for
other potential clumping factors or faint-end cut-offs to the LF using
Eq.~\ref{eq:convf}.

An escape fraction of $\sim$13-14\% would be much more consistent with
the low fraction of Lyman-continuum, ionizing photons confirmed to be
escaping from star-forming galaxies at $z\sim2$-4.  For example, work
by Vanzella et al.\ (2010) and Siana et al.\ (2015) estimate the
escape fraction to be $<$6\% and 7-9\%; meanwhile, analysis of the
afterglow spectra for a small sample of gamma-ray bursts (Chen et
al.\ 2007) suggest an escape fraction $f_{esc,rel}$ of 4$\pm$4\% at
$z\sim2$-4 (supposing $f_{esc,UV}$ to be $\sim0.5$).  While many
recent estimates of the escape fraction for $z\sim3$ galaxies yielded
values in the range 10-30\% (e.g., Nestor et al.\ 2013; Mostardi et
al.\ 2013; Cooke et al.\ 2014) and several apparent confirmations of
bona-fide Lyman-continuum photons (Vanzella et al.\ 2010; de Barros et
al.\ 2015; Mostardi et al.\ 2015), follow-up of many of the most
promising Lyman-continuum emitter candidates have shown that
foreground sources continue to act as a strong source of contamination
for such samples (Siana et al.\ 2015) despite apparently careful
efforts to accurately estimate the contamination rate using
simulations (Nestor et al.\ 2013; Mostardi et al.\ 2013).

Of course, in comparing the escape fraction estimates at $z\sim2$-3
with the inferred escape fractions at $z>6$ if galaxies drive the
reionization of the universe, we need to keep in mind the fact that
there must be some evolution in the escape fraction (e.g., Haardt \&
Madau et al.\ 2012; Kuhlen \& Faucher-Gigu{\`e}re 2012) to reconcile
constraints on the ionizing emissivity at $z=2$-6 (e.g., Bolton \&
Haehnelt 2007; Becker \& Bolton 2013) with the evolution observed in
the $UV$ luminosity density (e.g., Bouwens et al.\ 2015a).

\subsection{Implications for Escape Fraction Measurements with JWST}

In planning for future science endeavors with JWST, there is great
interest in devising strategies for measuring the Lyman-continuum
escape fraction from $z>6$ galaxies.  One possible approach for
measuring the escape fraction was proposed by Zackrisson et
al.\ (2013) and involved using various observed properties of a
stellar population, e.g., the observed $UV$-continuum slopes, to
predict the luminosity in various emission lines, particularly
H$\beta$, arising from that stellar population.  By comparing the
predicted luminosity with that expected from accurate stellar
population models, one could in principle infer the escape fraction
based on an observed deficit in the flux present in key emission lines
(i.e., H$\beta$ and sometimes H$\alpha$).

As discussed in \S3, there is some uncertainty in the present results
for $\xi_{ion}$ -- both because of the dependence on the dust law and
due to uncertainties on our stack results, i.e., $\pm$0.02-0.09 dex.
However, our results bring up an interesting possibility.  If current
estimates for $\xi_{ion}$ (adopting the SMC extinction law) are
correct and the intrinsic value for $\xi_{ion}$ is really in excess of
the expected values (based on $UV$-continuum information available for
galaxy samples) and the excess is $\sim$0.1 dex, this could be
problematic for the aforementioned strategy for measuring the
Lyman-continuum escape fraction.

The Zackrisson et al.\ (2013) strategy, while admittedly clever,
relies on our making accurate predictions for the overall output of
the Lyman-continuum ionizing photons from the continuum light produced
by stars.  If the escape fraction is not especially large (and 13\%
would appear to be an upper limit on its likely value), escaping LyC
photons would only impact the $H\beta$ luminosities at the $\sim$0.03
dex level.  If the intrinsic value for $\xi_{ion}$ cannot be
determined at the 0.02 dex level from the observations (much smaller
than the tentative $\sim$0.1 dex excess we find in $\xi_{ion}$ for SMC
dust), then it will be challenging to measure a positive escape
fraction at better than $2\sigma$ significance.

\section{Summary}

In this paper, we make use of a large sample of $z\sim4$-5 galaxies
for the purposes of estimating the Lyman-continuum photon production
efficiency $\xi_{ion}$.  Our selected sources were drawn from the
recent $z=3.8$-5.0 Smit et al.\ (2015b) and $z=5.1$-5.4 Rasappu et
al.\ (2015) selections, who make use of galaxies where the position of
the H$\alpha$ line in the IRAC filters with high confidence.  The flux
in the H$\alpha$ emission line is estimated by comparing the observed
flux in the $3.6\mu$m or $4.5\mu$m bands with the predicted flux in
this band based on an SED to the other photometric observations (see
Smit et al.\ 2015b; Rasappu et al.\ 2015; Shim et al.\ 2011; Stark et
al.\ 2013 for details).  We then use the inferred H$\alpha$ fluxes to
estimate the Lyman-continuum photon production efficiency $\xi_{ion}$
for galaxies in this sample.

In deriving the H$\alpha$ flux, we correct for dust extinction based
on the observed $UV$-continuum slopes while alternatively assuming a
Calzetti et al.\ (2000) and SMC extinction laws.  The H$\alpha$
emission line is assumed to be subject to the same dust extinction as
the stellar continuum.  We also suppose that 6.8\% and 9.5\% of the
flux at the position of the H$\alpha$ emission line is in the [NII]
and [SII] lines, based on both theoretical and observational results
for the line ratios (Anders \& Fritze-v.~Alvensleben 2003; Sanders et
al.\ 2015).

By applying this procedure to the $z\sim4$-5 galaxies in the Smit et
al.\ (2015b) and Rasappu et al.\ (2015) samples, we derive fiducial
values of $25.27_{-0.03}^{+0.03}$ and $25.34_{-0.02}^{+0.02}$ for
$\lxi$ assuming the Calzetti and SMC extinction laws, respectively.
The value of $\xi_{ion}$ for individual galaxies is estimated to show
an intrinsic scatter of $\sim$0.3 dex (Figure~\ref{fig:spread}).

This is the first time $\xi_{ion}$ has been estimated from the
inferred H$\alpha$ fluxes of $z\geq4$ galaxies.  $\sim$0.03-dex higher
values are expected if the escape fraction is non-zero and galaxies
provide the dominant contribution to the observed ionizing emissivity
at $z\sim4$-5.  The $\xi_{ion}$ values we derive would be higher
(0.02-0.09 dex) if we assume that nebular regions are subject to
$2.3\times$ higher extinction than the stellar continuum (as has been
found in the local universe: Calzetti et al.\ 1997).

The values we derive for $\xi_{ion}$ in the case of the Calzetti et
al.\ (2000) extinction law are quite consistent with canonical values
(Table~\ref{tab:xion_lit}) for all but the bluest sources.  For
sources with $\beta<-2.3$, we find $\xi_{ion}$ values which are
elevated by $\sim$0.2 dex relative to canonical values, as predicted
by various stellar population models considered in Bouwens et
al.\ (2015b) and Duncan \& Conselice (2015).

If the early ALMA results from Capak et al.\ (2015) are correct and
dust extinction follows more of an SMC-like dust law, the
Lyman-continuum photon production efficiency $\xi_{ion}$ we infer is
also consistent with canonical assumptions in reionization models, but
preferring a slightly higher value (by $\sim$0.1 dex) than in these
models and also as preferred for Calzetti et al.\ (2000) dust.

The high $\xi_{ion}$'s we measure for the bluest galaxies in our
selection are strikingly similar to those obtained by Stark et
al.\ (2015) on one $z=7.045$ galaxy, i.e.,
$\lxi=25.68_{-0.19}^{+0.27}$.  If these results are representative of
$z>6$ galaxies (and such results were suggested by work by Bouwens et
al.\ 2015b and Duncan \& Conselice 2015), this suggests that faint
blue galaxies may be especially efficient producers of ionizing
radiation.

The present results have important implications, implying that (1) the
stellar populations of $z>2$ galaxies produce Lyman-continuum ionizing
photons at the same rate as (or higher than) expected based on
standard stellar population models (e.g., Bruzual \& Charlot 2003) and
(2) indicating that galaxies cannot have an escape fraction
substantially higher than 13\% unless the $UV$ LF cuts off brightward
of $-$13 mag or the clumping factor is greater than 3 (\S4.2).  The
13\% escape fraction we refer to here is lower than the 20\% number
often used in conjunction with a Lyman-continuum photon production
efficiency of $\lxi=25.2$.  Unexpectedly high $\xi_{ion}$'s could be
problematic (\S4.3) for some proposed strategies for measuring the
escape fraction in the reionization epoch using the observed flux in
various recombination lines (e.g., Zackrisson et al.\ 2013).

We expect to extend these results to even lower luminosity galaxies in
future work.  This is by taking advantage of a larger sample of
$z\sim3.8$-5.4 sources with spectroscopic redshifts from MUSE (Bacon
et al.\ 2015) and the very deep 200-hour IRAC observations being
acquired over a 200 arcmin$^2$ region in the GOODS North and South
fields with the GOODS Re-ionization Era wide-Area Treasury from
Spitzer (GREATS, PI: Labb\'e) program (2014).  This same regime will
also be significantly explored leveraging the lensing amplification
achieved behind the Hubble Frontier Fields clusters (e.g., Coe et
al.\ 2015) as well as the deep Spitzer/IRAC observations and MUSE
redshift information available there.

\acknowledgements

This paper is much improved as a result of comments from an expert
referee.  We are grateful to Jarle Brinchmann for valuable
conversations concerning the H$\alpha$ fluxes in galaxies and the
robustness of conversions to ionizing photon production rates.  This
paper benefitted significantly from Brant Robertson's expert comments
generously provided pre-submission.  We also received valuable
feedback regarding current theoretical work in characterizing
high-mass stellar evolution (and the dependence on binarity, rotation,
etc.) from Selma de Mink and Ylva G{\"o}tberg.  We acknowledge useful
discussions with Jorryt Matthee and also support from NASA grant
NAG5-7697, NASA grant {\it HST}-GO-11563, and NWO vrij competitie
grant 600.065.140.11N211.

\end{document}